# A vision for global privacy bridges: Technical and legal measures for international data markets

*Sarah Spiekermann, Alexander Novotny* *

*Vienna University of Economics and Business, Department of Information Systems and Operations, Vienna, Austria*

ABSTRACT

From the early days of the information economy, personal data has been its most valuable asset. Despite data protection laws and an acknowledged right to privacy, trading personal information has become a business equated with "trading oil". Most of this business is done without the knowledge and active informed consent of the people. But as data breaches and abuses are made public through the media, consumers react. They become irritated about companies' data handling practices, lose trust, exercise political pressure and start to protect their privacy with the help of technical tools. As a result, companies' Internet business models that are based on personal data are unsettled. An open conflict is arising between business demands for data and a desire for privacy. As of 2015 no true answer is in sight of how to resolve this conflict. Technologists, economists and regulators are struggling to develop technical solutions and policies that meet businesses' demand for more data while still maintaining privacy. Yet, most of the proposed solutions fail to account for market complexity and provide no pathway to technological and legal implementation. They lack a bigger vision for data use *and* privacy. To break this vicious cycle, we propose and test such a vision of a personal information market *with* privacy. We accumulate technical and legal measures that have been proposed by technical and legal scholars over the past two decades. And out of this existing knowledge, we compose something new: a four-space market model for personal data.

"Only those who know the goal, find the way" Laotse (604 BC–531 BC).

## 1. Introduction

"Personal data is the new oil of the Internet and the new currency of the digital world" (2011). With these words Meglena Kunewa, Europe's prior commissioner in chief of consumer protection, expressed a current economic reality. Personal information (hereafter abbreviated as "PI") is at the core of online business models: it is regarded as the Holy Grail to gain the attention of users (Brynjolfsson and Oh, 2012). It also drives innovation because it promotes a better understanding of customer needs and reduces corporate cost and risks. Drawing on the analytics of Big PI Data, businesses try to avoid targeting the wrong customers, running into credit defaults, or hiring the wrong staff. Due to these benefits, the Boston Consulting Group predicts that the economic use of PI can deliver up to € 330 billion in annual economic benefit for organizations in Europe by 2020 (Bcg, 2012). The use of PI and its availability in emerging PI markets follow the predictions of neoclassical economic theory, which says that complete information and transparency creates economic efficiency (Posner, 1978).

* *Corresponding author*. Vienna University of Economics and Business, Department of Information Systems and Operations, Welthandelsplatz 1, D2, 1020 Vienna, Austria. Tel.: +43 1 31336 5279; fax: +43 1 31336 905279.
E-mail address: alexander.novotny@wu.ac.at (A. Novotny).

However, the strive for economic efficiency is challenged. People's legal right to keep their PI private and their right to informational self-determination (at least in Europe) limit marketers' push for the unrestricted flow and use of PI. Across the globe, regulators and human rights activists that see people's identities commercialized fight the “glass human being”. Across jurisdictions a delicate balance needs to be found between peoples' human right to privacy and data protection on one side and economic efficiency on the other (Wef, 2012).

To date, this balance has not been found. From a natural and legal point of view, only customer relationship holding institutions (hereafter abbreviated as “CR-H” for “Customer Relationship Holders”) should actually be entitled to use PI for the legitimate purpose of delivering agreed-upon services to customers. However, because PI is such an enticing tradable asset a plethora of data brokers have emerged to pursue more or less legitimate PI trade. An impressive “shadow market” (Conger et al., 2013, p. 406) for personal data has evolved that benefits from a current lack of global technical standards for controlling and auditing data flows. Nobody knows for sure who shares which PI with whom, in what form and on what occasions. Besides this nontransparent distribution structure of the PI market, data collection is mostly happening invisibly, without the full knowledge and true consent of customers (Angwin, 2012).

This status quo is unsatisfactory for all market actors involved. Customers and data protection authorities have reason to complain because they see the legal promise of informational self-determination and privacy increasingly eroded. For CR-Hs, the PI market lacks transparency and fails to create the necessary accountability and trust that is required for creating a predictive market environment. The markets' shadow existence undermines its own long-term viability, making all its players operate at the edge of what is ethically sustainable. A lack of accountability and transparency leads to an arbitrary valuation of the PI asset. And, as we will argue, it also fosters market concentration and impedes service innovation. Finally, different approaches to privacy regulation in the U.S., Europe and Asia lead to tensions that threaten data exchange.

Against this background, we strongly believe that a solid new vision is needed for how the PI market can work efficiently while providing privacy protection and informational self-determination. We don't think that global PI markets can simply muddle along with some technical and legal measures and compromises here and there. Instead we need a PI market, where

- companies handle PI transparently and accountably,
- long-term consumer trust is ensured,
- concentrated PI monopolies (or oligopolies) are broken up to make information more accessible to more companies on a global basis,
- and informational self-determination and privacy of customers is ensured.

If this is the goal, what legal measures would need to be taken? And what technical standards would be needed to enable these legal measures?

The contribution of this paper is that it provides a specific analysis how privacy-enhancing technologies and privacy/data protection law could together balance the right to privacy with market efficiency in order to create a trustful PI market. We present a vision of technical and legal bridges between continents that could be used by information systems and computer science researchers, standardization bodies and policy makers to challenge, streamline and prioritize current privacy regulation initiatives and technical developments. We believe that such a vision is a highly important tool, because what we need in this field is direction. As Laotse said once: “Only those who know the goal find the way”. Many aspects of the goal to have privacy-friendly digital services have been described already. Over two decades, legal and technical scholars in the field of data protection and privacy have dedicated their scientific lives to propose solutions to various aspects of privacy-friendly markets. Our aim in this article is to bundle the voices of these scholars into a chorus and to show how their solutions can be put together and integrated into one vision picture.

So far, only a few scholars have theorized about how a PI market vision could be organized with privacy in mind (Laudon, 1996; Noam, 1997; Schwartz, 2003; Aperjis and Huberman, 2012). These scholars have typically envisioned a market where people legally own and control their data, selling PI usage rights to data brokers under various organizational conditions. However, their proposed models fail to grasp the complexity of today's data handling practices; including the grown power structure of data markets. They hardly integrate the existing technological and legal landscape. And they provide no pathway to empower customers.

This article takes a different approach. We embrace existing work in the field of privacy research and outline how it could be leveraged in the current economic and technical environment. We show how privacy-enhancing technologies that allow for accountability (Pearson and Mont, 2011), agent-based privacy preference management (e.g., Cranor et al., 2006), cloud computing (e.g., Pearson and Charlesworth, 2009), anonymization (e.g., k-anonymity (Sweeney, 2002)), and differential privacy (Dwork, 2006) could be used to turn around today's adverse PI market situation. We think about incentives of current actors to participate in the market vision we propose. And we argue that only a symbiosis of “code and law” (Lessig, 2006) can produce an efficient PI market where customer rights to privacy are maintained.

To develop our PI market vision, we initially proposed a 3-tier market model (Novotny and Spiekermann, 2013) that categorized current PI market players into three groups: (1) CR-Hs involved in direct service- and PI exchange with customers, (2) data processing companies servicing CR-Hs and (3) third parties (including data brokers), which would work purely on anonymized data. This model was critically discussed and challenged in the course of 13 in-depth interviews with world-leading data protection experts (denoted hereafter as [I]) and a workshop with the European advertisement industry (denoted hereafter as [W]). The insights we gained through this empirical validation of our initial thoughts led to a new and final PI market vision that we present in this article.

To present our visionary market model, this article is structured as follows: In the next section, we give an overview of our market model vision. Then, we describe our methodology for elaborating and evaluating our PI market vision. We then describe our model in detail, outlining how experts have judged the pros and cons of our ideas. We summarize all of the legal and technical enablers that would be needed to put our vision into practice in table form. The paper closes with a critical discussion of our model's benefits and challenges.

## 2. The vision: a four-space market model for personal data and privacy in a nutshell

With his remark “Sapientis est ordinare” (Wisdom means to create order) Thomas Acquinas once hinted to the crucial power of order. To regain order, to create market transparency, trust and accountability, our vision is a PI market that contains four “spaces” and assigns standardized data handling and exchange rules to each of these spaces. We assign all PI market players to one space, though one company may take on different roles when they provide different services. Rules in each market space need to be enforced by technical infrastructure and regulation.

The first market space, which we call “customer relationship space” (1), includes customers and CR-Hs directly involved in a service exchange. For example, book buyer Bob may be a bookshop's customer, and the CR-H might be an online bookshop called bookshop.com.

The second market space, which we call “CR-H-*controlled* data space” (2), includes the distributed computing and service infrastructures that enables today's electronic business relationships. This space includes all companies providing services to the CR-H that *directly* enable and enrich the customer relationship. For example, a company in the CR-H-controlled data space could be the cloud service provider CloudServ, which handles purchase data for the CR-H bookshop.com.

The third market space, which we call “customer-controlled data space” (3), includes services that grant customers ownership of their PI and manage and control it in a privacy-friendly way. A service operating in the third market space could be MyShoppingBird, which collects book deals from various vendors, including bookshop.com under the control of buyer Bob.

The fourth market space, which we call “safe harbor for big data” (4), grants equal access to anonymized people data to all market entities that need it. For example, the market research agency NielsSearch could download aggregated people data from the safe big data space to analyze and forecast future consumer trends in the book market. Participants in this “safe harbor for big data” can provide and process as much data as they want, but the data they handle must be anonymized.

The safe harbor for big data is filled with data originating from PI. Yet, each time PI is transferred to the safe harbor for big data, it must pass an “anonymity frontier”. In fact, when PI leaves the context of an identified customer relationship and is transferred to an entity that is not involved in the customer relationship context, PI must cross the anonymity frontier. The anonymity frontier is operated according to “Best Available Techniques” (BATs) for anonymization.

Fig. 1 summarizes the four components of our market model vision plus the anonymity frontier.

Our market model vision has two major assumptions: First, we assume that societies can live with a “good enough” level of privacy protection for individuals. Our market model does not have a level of privacy protection that arms individuals against intentional, targeted attacks by organizations or criminals. We adopt the principle of proportionality that aims to balance economic utility with privacy (Iachello and Abowd, 2005). For instance, we embrace the fact that with access to voluminous data sets, anonymous data can theoretically be re-identified. Some criminal attackers may engage in such efforts even at the risk of criminal prosecution. But similar to locked cars and flats that can always be broken, we accept that anonymization techniques are “good enough” to thwart a lot of the crime potential.

Second, we assume that societies are ready to grant individuals the right to determine the fate of their PI. *Informational self-determination* can protect individuals from “information-rich perceptions by others” (Kaiser, 2012, p. 61). Control over personal information circulating in the market is at the heart of protecting privacy. In fact, privacy is often conceptualized as promoting control over one's personal information (Westin, 1968; Bélanger and Crossler, 2011, p. 1018) or as an ongoing process of controlling who has access to the self (Altman, 1977). Our market model therefore gives individuals the right to choose whether to participate in any collection, processing, or use of PI, and it recognizes that customers might reveal their personal data for benefits they value more than their data.

## 3. Method for evaluating the market vision

The interviews and the workshop to evaluate the market vision took place between October 2012 and March 2013. Interview partners included representatives from companies (including IBM, HP, Metro Group), standardization bodies (including the W3C DNT working group), data protection authorities (including ULD Schleswig–Holstein), data brokers (including a major credit-scoring agency), industry associations (including IAB), legal counseling groups and one NGO (EDRI). On average, interviewees had 11 years of work experience in the privacy domain. Although 12 of the 13 interview partners are based in Europe, eight are involved in global privacy efforts and working groups. Depending on their experience we made interviewees focus on at least one of the three "tiers" in our initial market model. The interviews and the workshop were audio taped and transcribed. The interviews lasted around 1 h, and the workshop lasted 2.5 h.

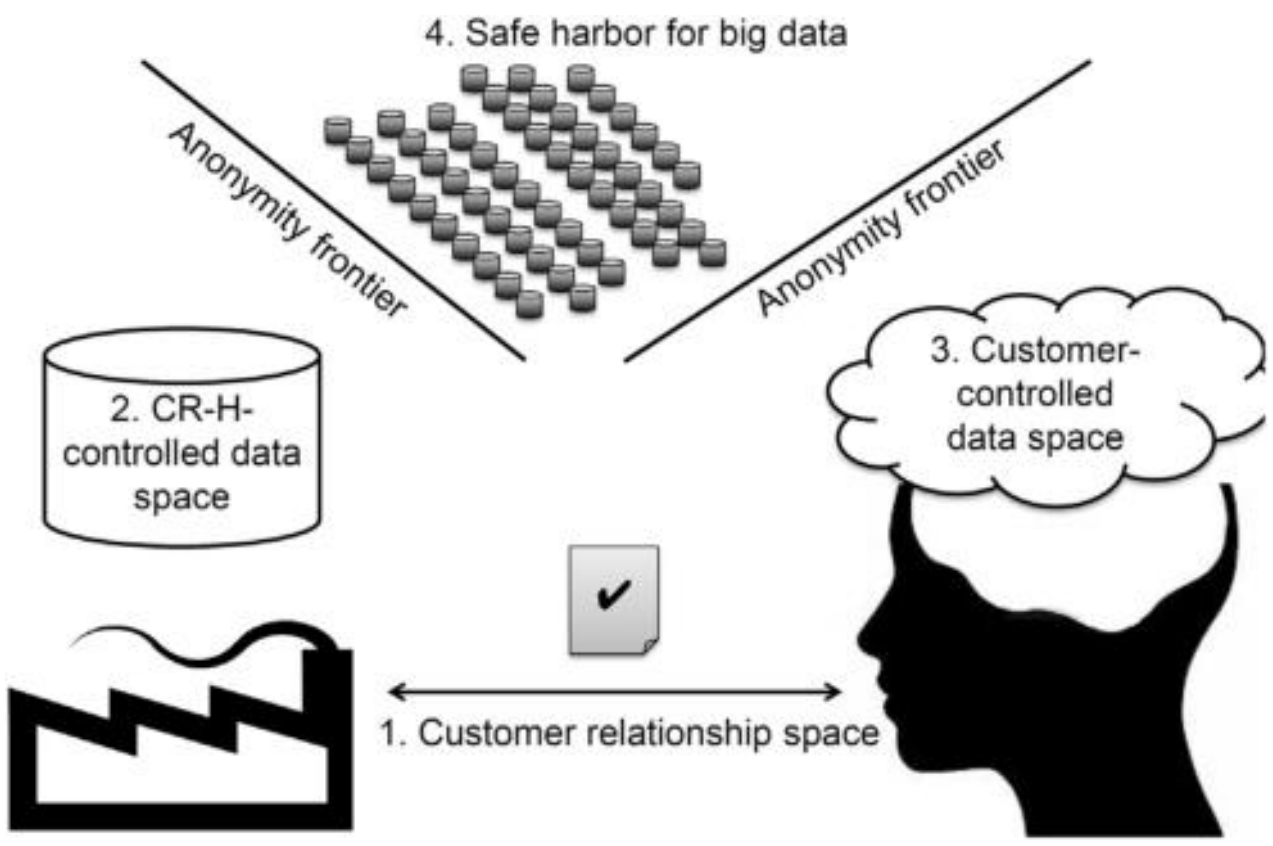

**Fig. 1 – Illustration of market model vision for personal data markets and privacy.**

Methodologically, we used a topic guide for interview consistency. After an initial question on interviewees' role and experience in the privacy field, we briefly introduced an initial version of the market model we had developed ourselves. This model is published in the Digital Enlightenment Yearbook 2013 (Novotny and Spiekermann, 2013). Interviewees were then asked to apply their own work experience to that market model. For example, we asked the W3C's DNT group representative to comment on our model's proposal to adopt agent-based (automated) privacy preference management for users to express consent; representatives from the credit rating agency were asked about the anonymization of data for brokering practices, etc. After this initially focused discussion, six core concepts of our market model were systematically evaluated in detail: (1) anonymization, (2) data use policies and their standardization, (3) technical accountability and auditability, (4) using and sharing anonymized information, (5) ownership rights in personal information, and (6) the transparent separation of service delivery from information exchange.

Interview transcriptions were analyzed for issues and arguments the experts raised on a topic (Froschauer and Lueger, 2003). First, in a within-interview analysis, passages in the text relating to one argument were paraphrased and summarized. Second, headlines were assigned to the paraphrased text passages; where possible, these were taken directly from text. Third, we compared all passages relating to one issue across the texts of different experts and the workshop, revealing the central arguments and core dimensions relating to each issue. Our analysis did not necessarily give preference to the interpretation of the majority of experts. Rather, disagreement between experts enriched arguments for and against our market model (Dorussen et al., 2005, p. 324). In a final step, we revised our initial model by incorporating the new ideas from the experts and outlined our final vision that is now going to be described in the next sections.

In the following sections, we outline the motivation, function, legal requirements, technical enablers and challenges in each of the four spaces. We also discuss the behavioral incentives of all market participants. We show that our market model encourages the development of innovative services based on PI and anonymized people data while ensuring a privacy-friendly use of customer's PI. Many of our arguments were finally supported and derived from the interviews we conducted; going beyond the initial market model we had come up with ourselves.

## 4. A critical discussion of the four-space market model and its components

In the following, we will discuss the four spaces of our revised market model in detail.

### 4.1. *The 1st market space: customer relationship space*

The core of the customer relationship space is the re-establishment of one-to-one business relationships. Companies that invest in a customer relationship need and want ***identified* customer relationships** (Spiekermann et al., 2003). And many customers are willing to provide their PI in a service context if it is necessary for service delivery or if they receive appropriate returns for their data. Therefore, our market model departs from the traditional data protection call for anonymity vis-à-vis *direct* CR-Hs (Gritzalis, 2004; Bella et al., 2011). Currently, an individual online customer often deals with multiple (n) parties collecting PI simultaneously during an electronic transaction with a CR-H. For instance, an average of 56 tracking companies monitor people's transactions on commercial news portals (Angwin, 2012). Thus, CR-Hs often serve as gateways to the shadow market of companies brokering PI. Our model requires that **only the one CR-H that is visible to the customer is allowed to collect PI** in the context of an exchange. This one-visible-partner rule is monitored by technical accountability-management platforms that are regularly audited and safeguarded by legal sanctions (see Section 6). Moreover, **CR-Hs become liable** for the proper handling of the PI they collect. All **PI they receive is recognized as being owned by the respective customer** and can be used by the CR-H only for purposes set down in electronic **PI usage policies**, which accompany each piece of PI exchanged. The PI and policy exchange are **automated with the help of privacy exchange protocols** (such as P3P (Cranor, 2012) or HTTP-A (Seneviratne, 2012)) that require minimal user configuration. These policies are the basis of a technically enabled and legally enforced **accountability scheme governing later PI exchange** in the CR-H-controlled data space. Finally, we suggest a **right to a privacy-friendly service version**, implying the **conscious separation of a service exchange from its information exchange**.

In the customer relationship space, we require identified one-to-one transactions and CR-H liability in order to ensure *predictability* of the outcomes of business transactions. Customers and CR-Hs are better able to appraise the costs of engaging in a business transaction when they have more information about the partner they are dealing with (Coase, 1960). Well-known transaction partners who send positive reputational signals are more likely awarded with *trust* in a business relationship (Puncheva, 2008). Especially in electronic environments, being able to see, perceive and feel a transaction partner increases the likelihood of engaging in a transaction (Aldiri et al., 2008). We re-establish predictability and trust in business transactions by forbidding third parties to invisibly observe customer relationships in the background. The latter practice is what causes many people to lose control over data collection. This again promotes distrust in electronic transactions and the CR-H (Smith et al., 1996; Spiekermann, 2005).

The following sections justify the characteristics of this space from an economic and human rights perspective and detail their technical and legal implementation (summarized in Table 1).

#### 4.1.1. *Principle of one-partner identification, visibility and liability*

Predictability is supported when users deal with only one visible PI-collecting CR-H at a time. We define *transaction partner visibility* as a state in which customers visiting an electronically enabled premise can unambiguously and effortlessly name the entity that they are transacting with; customers have an accurate mental model of who they are entrusting with their personal data. Transaction partner visibility is natural in the offline context. Customers visiting a physical retail store such as Walldepot perceive Walldepot as their CR-H. Likewise, in the online environment, when a user enters a bookseller portal such as bookshop.com, the visible partner brand is bookshop.com.

The concept of transaction partner visibility enables users to trust in the *contextual integrity* of their PI use (Nissenbaum, 2004). The contextual integrity of PI is preserved if disclosure and distribution of personal information adheres to socially accepted norms of appropriateness in a given social situation. Being able to identify one visible transaction partner is a social norm under which disclosure of personal data is appropriate. Users want identified relationships in which they can control the use of their PI. Contextual integrity is also a requirement for legitimate PI use. Any data exchange between customers and CR-Hs and any further PI use should be related to the transaction's purpose. Uses of collected PI unrelated to the transaction are illegitimate and the CR-H is held liable.

In contrast to what is common today, our market vision requires that the only entities entitled to collect or receive PI are those that invest in a one-to-one business relationship with a customer or that are assigned by the CR-H to help fulfill the contractual purposes. Today's data aggregators, trackers and brokers would need to invest in direct and visible service transactions with customers if they want to receive data. Alternatively, they can leverage their data processing knowledge and become players in the anonymous safe harbor for big data described below.

#### 4.1.2. *Ownership of personal information*

A core component of our model is that customers have a claim to *own* their PI, regardless of where it is held. The main reason for establishing an ownership claim to PI is a psychological one: “ownership” of PI creates asset awareness in the minds of all stakeholders. Customers who are aware that PI is an economic resource used to generate economic value make more informed decisions about disclosing PI (Spiekermann et al., 2012). Equally, companies will be more cautious and reflective in collecting and using PI if they are aware that they do not own the data; that the data is not their, but the customer's property. For these psychological reasons, we originally embraced the term “property right” to personal data as suggested by U.S. legal scholars in the field (Samuelson, 1999; Schwartz, 2003).

**Table 1 – Legal and technical enablers in the customer relationship space.**

| Legal requirements | Technical enablers: Mutual identity management standards | Standardized graphical user interface symbols for signaling PI usage policies and CR-holding companies | Software agent-supported PI usage policy negotiation and exchange | Client-side policy repository | PI usage policy languages with standard vocabulary |
|---|---|---|---|---|---|
| Mandatory ‘1-visible-partner’ rule | x | x | | | |
| Self-regulated standard on what is considered direct support of the CR to maintain contextual integrity | | x | | | |
| CR-holding companies obtain legitimization for PI use by:<br>• informed consent or<br>• legal empowerment | x | x | x | | x |
| Liability of CR-holding company for handling PI in accordance with electronic PI usage policies | x | | | x | x |
| Recognition of customers' PI ownership | | | | x | |
| Mandatory separation of the service deal from the PI deal | | x | x | | x |
| Obligation to offer one service option with minimum information use at reasonable quality and price | | x | x | | x |

Ownership claims are compatible with the human rights perspective of informational self-determination. The informed consent principle (Art 7 Directive 95/46/EC, Art 7 General Data Protection Regulation-draft (2012), Para. 7 OECD guidelines on the protection of privacy and transborder flows of personal data, Art 2 FTC Fair Information Practices (FIP)) and the right to object (Art 14 Directive 95/46/EC, Art 19 General Data Protection Regulation-draft (2012)) reflect that data subjects have the biggest interest in controlling the fate of their PI (Purtova, 2011). To recognize the human rights character of privacy, ownership claims to a data subject's PI cannot be alienated (Bergelson, 2003; Schwartz, 2003). In our model, only usage rights to PI can be traded for customers to better participate in personal data markets. This principle complements current data protection law and greatly facilitates customers' legal recourse (Buchner, 2011).

Our interviewees confirmed that customers *"intuitively have a right into their [personal] information anyways"[I4]* and recognized that the notion of ownership of one's personal information is simple and comprehensible. One expert argued that data ownership *"can be easily grasped by people" [I11]*.

Still, the majority of experts advised that the "ownership" of PI should be distinguished from private "property" rights. Many experts feared that viewing PI as property would not address the social complexity inherent in interpersonal relationships. A commercialization of individuals' identity seems to contradict the human rights perspective on privacy. Many experts therefore felt that a direct legal analogy with private property could be misleading. For instance, one expert pointed out that applying property laws to the use of PI by public sector institutions would imply an *"expropriation each time…"* and that would be *"an odd construction" [I12]*. Experts advised embracing the term "ownership" instead of "property right" *[I1, I7, I8, I10, I11]*. Viktor Mayer-Schönberger (2010, p. 1877) uses the term "quasi-property rights" to avoid confusion with property rights in a legal sense.

Regardless of experts' reservations, our proposal's ownership claims are compatible with the human rights perspective of informational self-determination. The informed consent principle (Art 7 Directive 95/46/EC, Art 7 General Data Protection Regulation-draft (2012), Para. 7 OECD guidelines on the protection of privacy and transborder flows of personal data, Art 2 FTC Fair Information Practices (FIP)) and the right to object (Art 14 Directive 95/46/EC, Art 19 General Data Protection Regulation-draft (2012)) reflect that data subjects have the biggest interest in controlling the fate of their PI (Purtova, 2011). To recognize the human rights character of privacy, ownership claims to a data subject's PI cannot be alienated (Bergelson, 2003; Schwartz, 2003). In our model, only usage rights to PI can be traded for customers to better participate in personal data markets. This principle complements current data protection law and greatly facilitates customers' legal recourse (Buchner, 2011).

#### 4.1.3. *Personal agent support of customers and legitimized PI use*

For the past 15 years, scholars have been working on privacy-enhancing technologies that facilitate (Mcdonald et al., 2009), automate (e.g., in scope of the PRIME Life project (Bussard et al., 2011)), and standardize (Cranor et al., 2013) the communication of privacy preferences. For example, people might object to data processing for credit scoring purposes or request deletion of their data after a period of time. Software agent (automated consent) solutions enable people to configure their privacy preferences in their IT clients (i.e., in the browser) and exchange these preferences with data controllers through PI policy exchange protocols. One of the most renowned software agent solutions has been the Platform for Privacy Preferences Project led by the W3C (P3p, 2006). Recent technical proposals going into the same direction is the HTTP-Accountability protocol proposed by Tim Berners Lee's group at MIT (Seneviratne, 2012) as well as the concept of "dynamic consent" developed for the UK health sector (Kaye et al., 2014).

Software agents facilitate the communication of privacy preferences to CR-Hs. Configuration effort is limited because the software agent learns the privacy preferences of the customer over time (Lodder and Voulon, 2002). The user's software agent compares PI usage preferences with companies' PI usage policies. The agent then signals to the user which service options or CR-Hs may be the most appropriate to engage with (similar to the 'Privacy Bird' presented in (Cranor et al., 2006)). It can also support negotiation of PI usage policies. The agent must ensure a voluntary, unambiguous, prior, verifiable customer agreement (Pachinger, 2012). It stores the usage policy and the contextual parameters of the customer relationship in which the policy is valid in a metadata-based architecture, such as the one proposed by Microsoft (Nguyen et al., 2013). The CR-H signals its consent and commitment to the PI usage policy to the user so that the user has a copy of the terms. The return-signal, potentially an extended version of today's DoNotTrack Feedback function, can act as a kind of receipt that records what PI has been provided to whom under what conditions.

By using a standardized, agent-based privacy preference exchange, standard PI usage policies are created for both customers and CR-Hs. These standardized PI usage policies are a basis for legitimized PI use in personal data markets. PI usage policies must be standardized. Standardized policies similar to the creative commons license types formulate the terms unambiguously and are easy to understand for consumers (Cranor, 2012). Moreover, they can be verified by customers when they view the settings of their personal agent (Art29wp, 2013).

The online advertising industry welcomed personal agent support of an informed consent, because this user-friendly way of giving consent seems to be the only way to make informed consent work. Today's manual consent to PI sharing is a too laborious and burdensome task *[W]*. One expert *[I9]* gave an example where a European DPA enforced cookie consent using modal pop-up windows on a company's website. As users had to consent each time before they were able to read the website, the number of users dropped by ninety percent. Seen such examples for manual consent, agent support was seen as a new avenue the industry could go. However, in order for personal agent-supported consent to be legally valid, one DPA-representative *[I12]* counter-argued that automating privacy-decisions would be problematic since consent would need to be given on a case-by-case basis. Configuring preferences beforehand and giving consent to all future cases would not be sufficient for the consent to be "informed". The compromise seems to be that personal agents need to nudge users into actively engaging with privacy settings *[I9]*. To fulfill the requirement of affirmative action, "*users need to be forced to actively engage with the settings of their agents and browsers.*" *[I9]*

That said, many experts were skeptical that personal agent supported PI usage policy negotiation would resolve the current "take-it-or-leave-it choice" that companies confront their customers with *[I8]*. The market asymmetry between CR-Hs and consumers would remain and undermine agent-based negotiation unless the legislator outlawed a coupling of services with data exchange. To alleviate market asymmetry, additional legal requirements, such as the right to a privacy-friendly service are therefore required as a complement (see below).

#### 4.1.4. *Separation of service and information exchange and the right to a privacy-friendly service*

Today, most online transactions are of a composite nature. Information is collected as a service spin-off (Jentzsch et al., 2012, p. 9) without making the 'information deal' visible to the customer. In our model, CR-Hs distinguish the service and information exchange within a business relationship. The service exchange includes the delivery of the principal service to the customer. For example, the principal service might be the sale and delivery of a book. Partners must offer one service option that requires customers to disclose only the minimum amount of PI necessary to fulfill the core service. Thus, people always have the right to a privacy-friendly service. Consider, for example, a web search engine. An individual opens the search engine's website, look.com. Selected by default, a privacy-friendly option may require the individual to pay a subscription fee of € 1 per month; this option neither records search queries nor shows any personalized ads. In contrast, the second option could cost € 1.50. This option collects more PI and only uses it for an agreed time period to provide a richer service experience, such as individualized search results (a better service compared to the first option's service is more expensive). The third option commercially leverages users' PI for an agreed time period for purposes such as the targeted placement of advertisements. This option may be provided for free. Users consciously trade their PI in exchange for the free search service.

The example demonstrates that the model would create awareness of the value of an online service on one side and the value of PI on the other side. Services won't necessarily come for free by default any more. The 'free' mentality governing online business relationships today would make room for a more realistic view of what digital services are worth. The advertising industry is convinced that if it is legitimate for a service to charge a fee, customers would be ready to pay *[W]*. Companies claim that personalized services enriched by PI provide value to the customer. If that claim is correct, customers should also be willing to pay for a better service quality, one expert argues *[I7]*.

But PI won't come for free either. Customers knowingly exchange their PI for a benefit, extending on strict data minimization. The separation of service options would improve the salience of the information transaction and encourage customers to make conscious decisions on information deals (Jentzsch et al., 2012, p. 10). Regulators believe that the separation of service and information exchange can help bringing more transparency into PI markets *[I7]*. Transparency is desperately needed, because customers are not aware that they pay with their PI in these markets *[I8]*.

#### 4.1.5. *Incentives of actors to embrace our customer relationship space proposal*

Customers are motivated to use personal agents to manage their privacy for several reasons: One is that they feel an increasing need to be protected from advertisements. 43% of online users use ad-blocking software today, a share expected to rise to 100% by 2018 (Pagefair, 2013). Another reason is that people's fear of losing privacy is increasing (Fujitsu, 2010), as is their computer literacy. Myriad companies have been founded in recent years that work towards customer-controlled architectures (Pdec, 2014). Software agents will improve the usability and information quality of informed consent procedures for customers. Instead of requiring customers to click checkboxes manually, they enable seamless rights management (Friedman et al., 2005). Moreover, if they empower people to manage the ownership of and usage rights to PI, they also technically enable customers to participate in personal data markets.

The one-partner rule enables CR-Hs to regain their rightful and exclusive control over PI collection in their transactions. Some CR-Hs, such as news portals and telecom operators, have lost this control in past years. Hence, the CR-H controlled customer relationship space will 're-decentralize' the collection and analysis of PI profiles in the hands of those who have nourished and built customer relationships. They should therefore be able to exclusively benefit from their customers' profile data. The control over data collection will also allow CR-Hs to re-establish accountability in their customer relationships, which is currently eroded by the multitude of parallel data collectors. Accountability is a vital requirement in Europe's new data protection legislation (Com, 2012).

CR-Hs should be willing to implement policy-based data exchange protocols because it facilitates consent procedures, automatically creates legal compliance for the data collected and may even improve the quality of data collected, as customers feel more comfortable entrusting them with timely and more complete data.

#### 4.1.6. *Challenges*

Market complexity may hinder the re-introduction of identified customer relationships based on the principle of the one-partner rule. In today's multi-entity service networks, it is difficult to distinguish between the different roles entities are taking on. "*Who is really the data controller…where am I, in service delivery, or at a third party?*" one legal advisory expert *[I9]* said, challenging our approach. Identification of the one company truly holding the CR-H may initially not be straightforward, though we believe that companies' market power will ultimately provide a clear definition. For sure, companies and the market as a whole would gain transparency regarding who actually owns the customer relationship.

Another challenge is for CR-Hs to draw the right line to maintain the contextual integrity of the PI they collect. Industries must establish standard rules for what is considered 'legitimized' use of PI in a context. Those companies operating in the "shadow market" will be challenged to transform their business models into legitimate ones. Respecting the one-visible-partner-rule, they will need to start providing services maintaining the contextual integrity of the PI they use.

Agreement also needs to be reached on what the 'necessary' minimum amount of PI is to fulfill a privacy-friendly core service. There is no simple formula for designing privacy-friendly services. Rather, trade-offs between functionality, usability, and processed PI have to be made on a case-by-case basis to ensure good-enough-privacy. The principle of proportionality has been applied successfully to make decisions of this kind (Iachello and Abowd, 2005).

Due to the numerous challenges that surround the well-established concept of informed consent, some have proposed to annihilate it completely. Critics of consent argue that obtaining it is practically burdensome, that customers are not truly informed, do not understand the consequences of giving consent, lack real choice in take-it-or-leave-it service offerings, or should not be able to trade away their data essential to their core of privacy. Others suggested that customers should not consent to the collection and processing of their data, but rather to the release of their comprehensive digital personas (Mpx, 2014). We believe though that we need not and must not give up on the informational self-determination of customers (i.e., control over data). The mechanisms our model proposes in the customer relationship space can support overcoming informed consent's burdensome handling; turning it instead into an effective instrument of customer control. As Edgar Whitley has argued alongside the UK's Technology Strategy Board, it is possible to give "consent as reliable and easy as turning on a tap and revoking that consent as reliable and easy as turning it off again." (Whitley, 2009). A combination of automated personal consent agents, one-partner visibility, true choice by obligatory privacy-friendly service offerings and ownership of personal data can facilitate such an effective, true and meaningful control for customers. Hard-law should mandate CR-Hs to obtain informed consent for legitimizing PI use not only under European legislation, but also in the U.S. and on an international basis.

Finally, PI usage policies may have scalability problems. Customers will not give separate consent for all transactions (Le Métayer and Monteleone, 2009), but their agents will need to provide appropriate responses to all consent situations. The agents will therefore need to learn and embed heuristics for disclosure decisions. They need to assure that agreed-upon PI usage policies are enforced within global service webs, which can be difficult given the number of shadow market players hungry for PI.

### 4.2. *The 2nd market space: CR-H controlled data space*

The CR-H is not the only party involved in the delivery of services and products. Subcontracting, outsourcing, and strategic alliances across multiple organizations are today's default. This complex service web of subcontractors receiving PI reduces the transparency and security of PI markets. Consumers are most concerned about secondary uses of their data by such invisible partners (Bélanger and Crossler, 2011). For this reason, we recognize the service web behind a CR-H as a market space that should be "controlled" by the CR-H, because the CR-H is that entity, which is finally seen as accountable for data breaches by data subjects. We therefore establish a 'CR-H-*controlled*' data space.

The CR-H-controlled data space comprises all those sub-contractors under control of the CR-H that directly contribute to the delivery of the CR-H's services. To ensure **contextual integrity**, sub-contractors contributions must be such that their receipt of PI can be anticipated by or justifiable to customers. Companies for which such contextual integrity cannot be justified don't qualify to receive PI. **CR-Hs are liable and accountable for their subcontractors**. For example, all application service providers that reach out to Facebook customers would be part of Facebook's (the CR-H's) controlled space. Facebook would become accountable and liable for any data breaches that occur within their partner network. Context-based trust between customers, CR-Hs and service providers are supported technically by **accountability management platforms**. These platforms manage the collection and sharing of PI based on the PI usage policies negotiated with customers. Through such platforms, accountability is created and authorization, non-repudiation, separation, and auditability of sharing practices are ensured. Legal safeguards must back up the appropriate use of an accountability management platform.

Technically enforced accountability re-establishes trust and transparency in personal data markets (Wef, 2012). Today's CR-Hs may have legitimate access to PI for specific purposes. But once they share their PI with third parties, strict compliance of that third party with the original terms of PI use is not guaranteed. Even from the CR-H's point of view, PI is flowing through a complex web of service providers that are difficult to keep track of. As a result, it is difficult to hold third-party service providers accountable. CR-Hs that collect PI and are responsible for keeping it under control are placed at a disadvantage. The following sections and Table 2 describe the technical enablers that support the legal requirements of creating the CR-H-controlled data space.

**Table 2 – Legal and technical enablers in the CR-H-controlled data space.**

| Legal requirements | Technical enablers: Use of an accountability management platform to enable and monitor policy-compliant use of PI | Technical enablers: PI usage policy languages with standard vocabulary |
|---|---|---|
| Liability of CR-holding company for handling PI in accordance with electronic PI usage policies | x | x |
| Obligation to implement and regularly audit the accountability management platform | x | |
| Separation of PI from multiple customers or CR-holding companies | x | |

#### 4.2.1. *Liability*

In our market model, the CR-H is legally liable for any collection and use of PI that violates the agreed PI usage policies. Liability arises for any violations by the CR-H itself or by any of the service providers under its control. Liability of the CR-H is natural from a customer perspective because the CR-H is the single point of contact for the customer. Its liability corresponds to customers' mental models. The interviews showed that claiming liability for data misuse in civic courts today is risky and expensive for customers because PI transactions are opaque. Legal experts, therefore, recommend a waiver of costs of litigation in data breach trials. *"Data has a value, but in the advocacy of my data I shall not pay legal charges." [I9]* Another privilege proposed by experts is shifting the burden of proof for proper data use (and lack of abuse) to the entity that benefits from using the data. Typically, this entity will be the CR-H. Our envisaged accountability management platform provides a technological basis for proving proper data use.

#### 4.2.2. *Accountability management platform*

To manage liability, the CR-H shall be responsible that all data use is traced on a technical accountability management platform. CR-Hs either implement the necessary accountability management platform by themselves or could use platforms operated by public entities in public-law data centers such as the German ZIVIT, for example. Technically enforced accountability ensures that any access, use, disclosure, alteration, and deletion of PI can be traced back to the originating party. Accountability management platforms must comply with the requirements of authorization, nonrepudiation, separation, and auditability. First, authorization requires that access to PI by the service provider is approved by the CR-H on an individual transaction basis. When a customer purchases a book, bookshop.com must explicitly authorize a credit-scoring agency to use the customer data to generate and provide the credit score of the customer. Second, nonrepudiation prevents service providers from falsely denying that they have accessed, used, altered or deleted PI. Any use of PI shall be traceable and recorded in logs, as proposed by those scholars who propagate auditability (Weitzner et al., 2008) and accountability (Seneviratne, 2012). Third, separation requires that PI originating from different service transactions, customers, and CR-Hs is strictly isolated unless the legitimate purpose allows for the combination of PI. This measure safeguards contextual integrity and thus the mental model of customers. Fourth, the platform should provide an audit trail. PI usage policies and rights attached to all data are used to demonstrate PI usage rights to authorities and auditors. They may also be used as evidence in liability processes. Early versions of such accountability management platforms are already under development and referred to as "trust frameworks" (Nguyen et al., 2013, p. 236) and "accountable HTTP (HTTPA)" (Seneviratne, 2012).

#### 4.2.3. *Incentives of actors*

CR-Hs have an incentive to implement an accountability management platform, because such a platform is an asset management tool in the end. The platform forces the service providers in the CR-H's network into managing their PI assets properly. Technical accountability brings intelligence and control over customer's PI home to CR-Hs. CR-Hs are the only parties that can analyze and use and thus benefit and monetize their customer's PI. Consequently, they should also bear the costs for operating the technical accountability management platform. An accountability management platform makes data flows and misuses of PI traceable. Technically assured traces of the PI flows allow CR-Hs to show that PI was handled in accordance with the agreed PI usage policy and the law. In case CR-Hs detect violations of PI usage policies, they can take redress against defecting service providers.

Technically enabled accountability management would also prohibit CR-Hs to act as a gateway to shadow entities. Privacy policies will clearly specify the one-visible-partner who can legitimately collect and use personal information from customers as well as aggregate profiles about them. Unauthorized shadow entities would be immediately detected on the accountability management platform automatically checking PI usage policies and reported. Some CR-Hs would be less forced into 'data-deals' with service partners.

#### 4.2.4. *Challenges*

Setting up and operating an accountability management platform is a highly complex and costly process. Multiple dimensions add to the complexity of guaranteeing traceability of PI on an individual transaction basis: the number of different organizational entities involved, the amount of PI transactions, and the number of automated interactions necessary between these organizational entities for a single transaction. One expert argued *"it is an additional complexity level...We do not talk about simply managing an additional set of data, but one has to check for each handling of personal data whether one is still allowed to access this data."[I3]* This complexity is additionally driven by processing personal data in distributed or remote systems, such as in cloud systems. Often, large companies have no consolidated overview of data flows, where this data is stored, and who has access to it *[I9]*. Thus, for some companies it may imply a high effort and heavy investments to build up an accountability management system.

### 4.3. *The 3rd market space: customer-controlled data space*

Privacy scholars and some start-up companies have suggested relocating personal data into a sphere solely controllable by customers. They have proposed identity management platforms (e.g., PRIME and PRIMELife Project) that help customers to **manage their PI** and the various identities linked to them. The platforms act as **intermediaries for customers**, potentially pulling services for them from the market. They take notes of what customers have revealed to whom. In a more sophisticated scenario, they mine personal activities and begin to understand customers' preferences based on user-side data mining (Lodder and Voulon, 2002). These preferences are then used to support customers in their online activities such as the search for product offerings. The customer-controlled data space is enabled by **trusted third parties** and **personal data vaults**. Compared to the CR-H-controlled data space, these offer customers increased control over PI storage location, intelligence applied to PI, and PI deletion. Table 3 gives an overview of the legal requirements and technical enablers proposed for such a customer-controlled data space.

We have added a customer-controlled data space to our original model version, because it will play an important role in the future for privacy preservation. In fact, identity management systems have recently evolved to include personal data vault technologies, such as user-centered social networks (Baden et al., 2009). In an entrepreneurial effort in Silicon Valley, Kaliya Hamlin, who calls herself "Identity Woman", has established the Personal Data Ecosystem (Pdec, 2014), which supports small companies with venture capital access and knowledge to pursue a user-centric data-handling strategy. In an extremely visionary way, Doc Searls proposed the establishment of an "Intention Economy" (Mitchell et al., 2008) where customers use personal agents or third parties to pull services from companies rather than companies proactively approaching customers and offering their services. Trusted third parties and personal data vault technologies are key for such ideas to thrive.

#### 4.3.1. *Trusted third parties*

Trusted third parties manage personal data on behalf of customers and mediate business relationships between customers, CR-Hs and the safe harbor for big data. These third parties act exclusively on behalf of customers and aim to give them full control over their data. For example, InternetPost AG's safe address platform manages personal data that is used by CR-H's on behalf of the customer (Internetpost, 2014). Trusted third parties provide customers with client software and possibly hardware that collects PI and securely transfers it to the third party's repository. Customers can access, modify, or delete their PI in that repository. They may grant CR-Hs access to parts of their PI, which can be used to enhance a business relationship. Personal agent technology operates on top of a customer's data vault and third party infrastructure creating a seamless PI flow from customers through third parties to CR-Hs or the safe harbor for big data. Personal agents may be provided to customers by their trusted third party.

#### 4.3.2. *Personal data vaults*

If customers do not want to store all of their PI with a third party, they can use personal data vaults. Personal data vaults are storage systems for PI which they operate themselves (Mun et al., 2010). They can be as simple as plug computers that are always online. The PI stored in data vaults can be accessed by standardized protocols that enable interoperability between the personal data vault of a customer and the CR-H, potentially mediated by a third party (Narayanan et al., 2012).

Personal data vaults enable customers to collect, store, analyze, change, or delete PI without any other party being involved.

#### 4.3.3. *Incentives of actors*

For customers and CR-Hs, customer-controlled services offer several advantages. Customers benefit from *increased privacy* because they maintain tighter control over PI handling practices. Both in the trusted third party and or personal data vault models customers' informational self-determination is strengthened. They can decide where data is stored (storage control), how it is analyzed (aggregation control), when it is deleted (retention control) and how much is revealed (disclosure control). When service providers interact with customers in the customer-controlled data space, the customer, through his or her trusted third party or personal data vault, is the nexus of PI exchange. As a result, we presume that the network of service providers through which personal data flows will become less complex than it is today. Clear perceptions of PI handling practices make users aware of these practices and increase their perception of control (Böhme, 2009). Furthermore, with PI handling practices under their full control, users may *build more trust* in the parties handling their data.

**Table 3 – Legal and technical enablers in the customer-controlled data space.**

| Legal requirements | Technical enablers: Advanced crypto-protocols for a secure exchange between user clients and trusted third parties | Mutual identity assurance standards | Software agent technology: Highly usable customer control over settings, ability to learn privacy preferences | Easy-to-use, user-operated personal data vaults | Policy languages for data exchange between customers and companies |
|---|---|---|---|---|---|
| Obligation of trusted third parties to act on behalf of the customer | | x | x | | |
| Obligation of trusted third parties to keep PI of customers secure | x | x | | | |
| Liability of CR-holding company for handling PI in accordance with electronic PI usage policies | | | | x | x |

CR-Hs and their service providers benefit from *lower market-entry barriers*, a higher legal certainty and higher data quality. In the CR-H-controlled data space, seen the established competition today, it is difficult for new market entrants to scale up because gaining access to a mass of customers from which data can be collected is difficult. Without enough data, companies cannot provide users with the same service experience as established players. In fact, most CR-Hs fail to achieve a critical customer base allowing them to build positive network effects (Shapiro and Varian, 1998). In our customer-controlled data space, companies do not need to obtain such a critical mass of customers; they can access and obtain behavioral data from the safe harbor for big data. Take the start-up of a new navigation and traffic service company as an example. Today traffic data is monopolized by a few big players, such as Google, Apple or TomTom. The start-up has no chance to provide a comparable service because it does not have access to real-time traffic data. In our model anonymized traffic data is shared among all players. The start-up can immediately start its new service.

CR-Hs also benefit from *higher legal certainty because of better predictable liability*. CR-Hs liability is better predictable, because they do not remain accountable for a complex and hard to control network of service providers. Their liability is restricted to have an accountability management platform in place enabling them to proof their lawful own use of the data. Rather than being in the role of data controllers, CR-Hs are just temporary users of the PI stored at trusted third parties or in personal data vaults that customers provide access to. Third, CR-Hs benefit from *higher data quality*. As the data is stored and maintained under customer control at trusted third parties or in personal data vaults, companies get data firsthand. The risk of operating on "dirty data", data which is wrong, outdated, or inaccurate, is reduced (Strong et al., 1997; Kim et al., 2003).

The interview results support the idea that trusted third party services may be a viable business model. Experts argue that managing customer preferences is a business model different from targeting CR-H offers to customers, but current advertising networks have the operational knowhow to migrate into such a business *[W]*. Trusted third parties could finance their operations from a commission imposed on successful leads mediated between customers and CR-Hs. 'Soft IDs', user pseudonyms issued by the trusted third party in the course of a transaction, may also allow trusted third parties to take over the role of financial- or identity brokers *[W]*.

#### 4.3.4. *Challenges*

The customer-controlled data space counts on the *emancipation* of customers and on their time investment in the management of their PI. Customers with a high internal locus of control (Xu, 2010) will be at an advantage to capitalize on informational self-determination. Customers must make a number of decisions on their own: selecting a trusted third party, deciding on operating a personal data vault, or determining when PI shall be deleted. One expert highlights that research is required on whether *"the market is actually ready or whether it is only a niche model for the enlightened"[I13]*. Currently available *personal data vaults* are *difficult to use*. They require custom configuration and the installation of additional software and hardware (Narayanan et al., 2012). Effort has to be made to get customer-controlled data technologies to reach a wider audience than a few select innovators.

Trusted third parties can be a *single point of failure* because they bundle the trust of customers and the risk of storing extensive electronic identities. Regarding security, third parties must consider not only the storage of PI, but also how PI can be transferred to and accessed by them. One prerequisite is therefore *secure interfaces to trusted third parties*, advanced crypto-protocols that allow for a secure exchange of PI between user clients and trusted third parties. Experts point out that these interfaces need to be trusted and prevent impersonation; it would be necessary to *"get to a point where one can authenticate electronically without doubt" [I10]*.

### 4.4. *The 4th market space: safe harbor for big data*

A core component of our market model is the safe harbor for big data. This part of the market contains anonymized PI for all players. We use the term '**people data**' for anonymized PI. Customers can voluntarily 'donate' their data to the safe harbor for big data. For example, individuals may share their navigation patterns with the safe harbor for big data so anyone can benefit from traffic congestion information. CR-Hs and trusted third parties may transfer data to the safe harbor for big data on behalf of their customers. Each time PI is transferred to the safe harbor for big data, it must cross the '**anonymity frontier**'. Based on a **principle of reciprocity**, everyone has equal access to the data stored in safe harbor for big data. Entities may only draw data from the space if they contribute proportionally to it.

#### 4.4.1. *Open people data and reciprocity*

People data (anonymized personal data) in the safe harbor could be considered a public commons. Similar to a creative commons zero (CC0) license, the purpose of the use of people data cannot be restricted. Data may be used for commercial or noncommercial purposes without providing attribution, and new data may be derived. Open people data does not imply that it is free. Providers may host the infrastructure of a safe harbor for big data, charging a nondiscriminatory access fee to cover operational costs. The safe harbor should foresee the reciprocity of data exchange: entities are allowed to use data from the safe harbor for big data only when they contribute to the space proportionally. Proportional contribution may not be understood in terms of data volume, but in terms of an entity's individual ability to contribute. And no entity should be excluded from gaining access if the entity fulfills the requirement of reciprocity. The following paragraphs discuss the legal requirements and technical enablers to build a safe harbor for big data (Table 4).

#### 4.4.2. *Anonymity frontier*

The safe harbor for big data is a privacy commons, a shared space of anonymity. When data enters the safe harbor for big data, it crosses the boundaries of a customer relationship's contextual integrity. PI therefore has to be anonymized at the anonymity frontier. Anonymization would need to occur according to 'best available techniques' (BATs) for anonymization (see Section 4.5.1 below for more detail). The latest technical anonymization concepts such as k-anonymity (Sweeney, 2002), l-diversity (Machanavajjhala et al., 2007) and t-closeness (Li et al., 2007) would need to be combined and regularly reviewed by a global group of experts to ensure that data passing the anonymity frontier sees indeed a strong degree of data generalization. The BATs could include such guidelines as a k-anonymity-threshold level (for instance a threshold of $k = 100$), meaning that only those entities can contribute people data that can allow for this degree of anonymization. This means that PI in some contexts may *not* be viable to be anonymized at all. For instance, PI from genetic research cannot be viably anonymized, because an individual's footprint is so unique that the number of people sharing the same genetic characteristics is effectively zero ($k = 1$). Depending on industry-specific privacy risk assessments certain domains would therefore need to be excluded from distributing PI to the safe harbor for big data.

**Table 4 – Legal and technical enablers in the safe harbor for big data.**

| Legal enablers | Publicly accessible infrastructure hosting safe big data (Technical enablers) | Anonymization technology used at anonymity frontier (Technical enablers) | Accountability management platform monitoring policy-compliant use of PI (Technical enablers) |
|---|---|---|---|
| Open, reciprocal access to safe big data | x | | |
| Legal anonymity requirement in ‘safe harbor for big data’ | | x | x |
| Sanctions for violation of anonymity requirement | | | x |

Creating such a ‘safe harbor for big data’ requires courage of policy makers to deviate from how data protection authorities want to restrict the use of anonymized data today. Up to now, the UK's Information Commissioner's Office interpreted the EU data protection directive that “consent is generally not needed to legitimise an anonymisation” (Ico, 2012, p. 28). In 2014, however, the Art. 29 Working Party published its opinion that even for anonymized data the purpose limitation principle would need to be followed and a grounds for legitimate processing (e.g., informed consent) would be required (Art29WP, 2014, p. 7). Seen that there are no globally accepted BATs for anonymization, this position seems justified. There is no rigorous ‘anonymity frontier’ as we envision it here (see Section 4.5) and hence there is a risk that data is being circulated and used as ‘anonymous’ which is easily identifiable. There is also no criminal prosecution for re-identification. Against this background the Art. 29 WP is wisely suggesting to apply the same rules of caution to data that is called ‘anonymous’ as to identified data. As we will outline below, though, our approach to anonymization is very rigorous from a technical and legal standpoint. And against this background we argue in line with the UK's ICO (Ico, 2012) that the positive societal benefits that could arise from open people data should be leveraged.

#### 4.4.3. *Incentives of actors*

The main motivation to introduce a safe harbor for big data is innovation. Across all industries, data-driven companies tend to be robustly more productive and more successful than their low-volume data competitors (Mcafee and Brynjolfsson, 2012). Big data has the potential to improve the quality of decisions in all areas of life, where previously only small sample sizes or exclusively qualitative arguments have been considered (Mayer-Schönberger and Cukier, 2013). For example, big anonymous people data was used to predict the spread of malaria in Kenya, to analyze population migration after the Haitian earthquake in 2010 and to understand the socio-economic development of communities in the UK (Unglobalpulse, 2013). Currently, however, truly big people data is only available to a few major corporations, most of which are based in the U.S. (i.e., Google, Apple, Facebook, Amazon, major national portals). These companies can leverage their customer's data to re-enforce their quasi-monopolistic market positions. Smaller companies, in contrast, have difficulty ramping up innovative services because they cannot access a database of user behavior, which is often needed for service design. For example, offering an innovative traffic jam warning system is difficult if congestion information is oligopolized by Apple's, Google's and TomTom's map services. Granting access to people data in the safe harbor for big data on equal terms to all market players globally could *reduce winner-take-all dynamics* in data markets generally. Market-entry barriers would be reduced for smaller companies, paving the way for vital competition (Shapiro and Varian, 1998). Monopolization of insights into social behavior will be limited. One effect could then be that the competitive edge of companies will shift from amassing PI to competition based on characteristics of service quality. A company's success will not depend on *having* the most data any more, but being able to make the best sense out of the anonymized people data for the user. Competition based on intelligence, functionality, and usability is stimulated. Consequently, customers will benefit from additional value extracted from anonymized people data while maintaining privacy.

Data donations to the safe harbor for big data may be driven by *philanthropy and altruism* as well as the desire to improve service quality. In fact, people already donate data. For instance, users send reports on crashed applications and usage patterns to application providers for fixing bugs or improving application quality. “Identity woman” Kaliya Hamlin calls this practice “data raindrops”.

#### 4.4.4. *Challenges*

There is no such thing as perfect anonymity. By now it is clear that with sufficient third-party data, money and motivation there is a residual privacy risk when malicious attackers want to re-identify data donators (Ohm, 2010; Bambauer et al., 2014). We recognize and embrace this threat. Just as with physical locks that we use to close apartments and cars we believe that anonymization is a 'good-enough lock' for ensuring a high level of privacy for a majority of people. We do not deny that the door may be forced open, but we argue that despite the residual risk of re-identification the level of privacy would be very high in a safe-harbor for anonymity. This is, first, because re-identification is legally penalized in our model. It is criminally pursued and prosecuted (see Section 4.5.2). And second, the protection of people data can also be technically supported: information downloaded from the safe harbor for big data can be watermarked with an irremovable sticky policy for instance, making re-identified information immediately visible to auditors and authorities. With such watermarking we would also prohibit what Paul Ohm has called a "release-and-forget" practice once data is anonymized (Ohm, 2010, p. 1711). "Release-and-forget" means that data is made public without ever tracing the further purposes of its use.

**Table 5 – Legal and technical enablers of the anonymity frontier.**

| Legal enablers | Anonymization technology | Accountability management platform monitoring policy-compliant use of PI |
|---|---|---|
| | Technical enablers | |
| Best available techniques reference documents defining best available anonymity | x | |
| Auditing of adherence to best available anonymization techniques | | x |
| Sanctions for violation of anonymity requirement | | x |
| Right to freely exchange information with standardized anonymity | x | |

Moreover, we want to make a remark on the final risks for people caused by re-identification. This risk includes the amount of psychological, financial or physical harm that would be caused by an act of re-identification. The original motivation of data protection regulation is to minimize serious risks for people; to not get hurt. But in many cases re-identification of data does not cause harm to people. The humanitarian and societal consequences of seeing some instances of re-identification would probably not be devastating. As Ohm (2010) argues, policy makers need to move away from anonymization math to sociology. What counts is the potential risk for people after people data's re-identification, not the act of re-identification itself.

Besides those arguments on the technical limitations of anonymity, an economic challenge has been brought forward. One was that there could be an imbalance between those who contribute data and those who profit (as is the case in today's data markets). The practice of downloading data from the harbor without contributing to it could be argued to threaten the quality of this big data pool in the long run. However, in our model vision, the safe harbor for big data discourages this practice by allowing access to data only on the basis of *reciprocity*. Fairness is guaranteed by requiring proportional contribution to the space when data is consumed. Even when there are no perfect reciprocal contributions, evidence from open source software communities shows that similar exchange systems work when the number of contributors is skewed (Lessig, 2006).

While the safe harbor for big data offers a lot of open data, its user might not be able to extract a lot of information from it. Because the data is available to everyone, the economic *value of the information approaches zero over time*. Data posted to the safe harbor for big data evolves from a common good that people compete for to a shared public good over time. Also poor analytical capabilities could be the reason. Companies may give priority to statistical correlations turning up in the vast amount of data instead of analyzing meaningful causal relationships providing utility to service users (Mayer-Schönberger and Cukier, 2013). "It's easier to get the data in than out" (Jacobs, 2009, p. 69).

Moreover, a tradeoff exists between fragmentation of the safe harbor for big data and its scalability. On the one hand, a fragmented safe harbor for big data is consisting of smaller, open repositories that focus on a narrow range of data within a domain. The utility of fragmented spaces would be limited. But they would be easier to deploy and be adopted by market participants more quickly. On the other hand, a comprehensive space containing a higher volume of different types of data from versatile application domains would provide high utility, but it may not be scalable and involve higher risks of unauthorized re-identification.

## 4.5. *Standardized anonymity frontier*

The anonymity frontier separates identified, personal data from non-identifiable, anonymous information in the market. PI crossing the anonymity frontier has to be sufficiently anonymized so that re-identification is not "likely reasonable".[1] What constitutes sufficient anonymity is defined in a generally accepted **standard for anonymization based on timely best-available techniques** (BATs) for anonymization. Adhering to BATs, engineers and operators of anonymizing information systems need to ensure that information meets a reasonably safe level of anonymity using best-available techniques. Possibly operators may automatically check the anonymity level by support of algorithms. Reasonably anonymized information may be shared and freely exchanged without restrictions. To secure customer's privacy at the anonymity frontier, **heavy legal sanctions** need to be imposed on those who do not respect it (Table 5).

The benefits of a standardized anonymity frontier are manifold: First, standardized anonymity will shift competition on data access to other characteristics of a service (e.g., quality). There shall be *no competition on privacy*. From a fundamental rights perspective, all individuals have an inalienable core of personal privacy regardless of their socioeconomic status or market power.[2] To guarantee a universal minimum level of what constitutes non-identified use of information outside of privacy boundaries, standards for reasonable anonymity are required. Standardized anonymity also *reduces the transaction costs* of companies that use data. Companies save on the costs of defining and individually negotiating what constitutes reasonable anonymity (e.g., required size of the anonymity set (Sweeney, 2002)) for each type of transaction. Rather, they can rely on collectively accepted standards of anonymization for a given type of transaction. For example, the U.S. Federal Trade Commission clarified the standard of "reasonable linkability" to provide companies with a commonly accepted interpretation (Ftc, 2012, p. 18). Accepted standards expose them to *smaller risks for damage to their reputation*. Companies affected by a breach of anonymization can defend accusations of misconduct and avoid liability by proving that state-of-the-art anonymity standards were followed. Also, the risks of misguided investment in unacknowledged anonymization technology are reduced. Finally, the unrestricted use of de-identified data is facilitated for companies. Companies often use aggregated data relating to a group of individuals. The standardized anonymity frontier allows companies to share anonymous data with *any* other entity, in-house or outside of the company, without user consent or adhering to data protection regulations. Even though data protection regulations will not be applicable to anonymous data in our model, companies still need to obey privacy laws,[3] for example to prevent that unfair automatic decisions are taken based on anonymous people data.

[1] See Recital 26 of directive 95/46/EC provisioning that "account should be taken of all the means likely reasonably to be used either by the controller or by any other person to identify the said person".

#### 4.5.1. *Best available anonymity*

What constitutes sufficient anonymization is a dynamic concept dependent on the state-of-the-art of technology. Regulators should document and update current standards for anonymization in so called "BREF"s, best available techniques reference documents (IPPC, Directive 2010/75/EU). The definition of BATs for anonymity in BREFs follows procedures of global co-regulation in which recognized and trusted experts take part. Currently, the concepts of k-anonymity (Sweeney, 2002), l-diversity (Machanavajjhala et al., 2007) and t-closeness (Li et al., 2007) suggest that it is sufficient to have a large anonymity set of individuals, diverse attribute values and similar attribute value distributions. The fact that best available anonymity techniques can be successfully defined has been demonstrated by the processing of customer health data in German pharmacists' information system (Giesen and Schnoor, 2013).

Because of their brisance for the protection-level of the anonymity frontier, the definition of best available anonymity standards need to follow a transparent process. Public representatives (e.g., NGOs such as EFF[4] or EDRi[5]) should be able to participate in writing down BREFs for anonymity. This high level of transparency and participation can be achieved by holding BAT standardization meetings in public and running through a request for comment (RFC) process[6] after initial draft rounds. Also, to incorporate advances in re-identification technology and computing power, BAT standards can be regularly updated, for example following a cycle of every three years.

The experts highlighted the relativity of the anonymity concept pointing out rightfully that there is no absolute anonymity. "You don't actually need that much external information to add to the information you have got to allow you *identification*" one expert *[I1]* highlighted. Virtually any large data set having a high entropy can be de-anonymized; a U.S. inhabitant, for instance, from which only ZIP code, date of birth and gender is known can be identified with a probability of 87% by his or her real name (Sweeney, 2002). And with the big data trend, entropy in data sets is rising swiftly (Rubinstein, 2012). Despite these mathematical limitations of anonymization (Ohm, 2010), experts are convinced that anonymization *"will be essence in the future" [I3]*. From practical experience, experts report that the k-anonymity-threshold level can often be increased with low technical effort *[I9]*. For instance, for many applications only the year of birth, not the exact date of birth needs to be recorded. Another misguided argument against anonymization is that individual characteristics are known with certainty. Consider a dataset in which 20 attributes are known with 80% probability. Because probabilities multiply for all attributes, there is only a 1.15% chance (0.8 to the power of 20 attributes) that the profile accurately describes an individual. In online advertising user interest profiles, for instance, attributes are almost never known with certainty and profiles are incomplete with many data points missing *[W]*. "Often the approach [to anonymization] is too much Newton, but we have to go for Heisenberg" *[W]*.

#### 4.5.2. *Sanctions*

To guarantee that best available anonymity standards are obeyed and can be effectively enforced, penalties and damages for the illegal acquisition, possession, use or sale of identifiable information are necessary. Sanctions protect a trustworthy market regime. Therefore, sanctions are required even in case no harm was caused. If any entity outside the contextual boundaries of a customer relationship is caught engaging in PI storage or processing to which it is not entitled (because no authorization policies can be provided), it shall not only pay damages to customers, CR-Hs and others but also pay substantial punitive damages (Traung, 2012, p. 42). Moreover, any natural persons involved in the illegal activities shall face criminal prosecution, as they have encroached upon the fundamental rights of other individuals.

[2] See Art. 8 of the Charter of Fundamental Rights of the European Union.

[3] In contrast to data protection regulation, the Art. 29 Working party names the e-Privacy directive, Art. 8 ECHR and Art. 7 of the EU charter of Fundamental Rights as example (Art29WP, 2014).

[4] Electronic Frontier Foundation.

[5] European Digital Rights initiative.

[6] RFC processes are well-known participatory community processes in the Internet-domain and have been successfully adopted by the Internet Engineering Task Force (IETF) to develop and norm foundational technical standards for the Internet.

#### 4.5.3. *Incentives of actors*

Customers and data-using companies have *opposing incentives* as to which standard for anonymity shall be defined in best available techniques reference documents. While consumers favor strong definitions of anonymity, data-using companies prefer weak anonymization technologies and administrative measures. *"The anonymity frontier shifts by the best available techniques standard; there is an enormous pressure to improve this standard while there are economic incentives to research less on anonymity."[I6]* one expert characterized the situation. For such a system to work, therefore, roughly equal powers must come to the negotiating table. Consumer and privacy pressure groups as well as industry representatives need to reach a balanced agreement on what is deemed sufficient anonymization. Anonymity BREFs need to strike a balance between privacy and data utility.

Some companies may also try *undercutting* what is defined as the *minimum standard of anonymity*. *"The more comprehensive the data, the more valuable it is"* one expert *[I6]* concluded. Some companies outside of the contextual boundaries of a customer relationship may be encouraged to re-identify data. But the marginal utility from identification outside of such relationships is likely to be so minimal that the ensuing risks of getting caught are not justified. Severe *sanctions* also deter this illegal behavior. Laws can be effectively enforced by tracing the audit trail of PI recorded in the accountability management platform that customer-relationship companies need to implement.

#### 4.5.4. *Challenges*

High *security* requirements and trust is put into the points responsible for anonymization. Anonymization itself has to be a minimum vulnerability procedure: *"high trust is required into this anonymity services, because, unfortunately, it occurs that such data is stolen frequently…Whoever gets certified… [by a best available anonymity standard] should ensure that the data are not hackable" [I2]* one expert says. A company's information security management system has to recognize that interfaces at the anonymity frontier represent vulnerable targets. Therefore, one expert argues, one "*needs administrative controls*", because *"purely technical controls are necessary, but in no case sufficient." [I6]*.

Another challenge is drawing the exact line of the *contextual border within a group of companies* or possibly the company itself. One industry expert of a customer relationship holding company *[I4]* pointed out that "*for instance, there is the procurement department…These [the data] are absolutely anonymized*." Especially the corporate headquarters often would not accept that identified customer data of subsidiaries cannot be used for other purposes in other companies a legal expert highlighted. Large international organizations may therefore have to invest substantial effort to disentangle and elucidate internal flows of PI.

## 5. Leveraging existing research for the four-space vision

Technical enablers support the enforcement of our model's legal requirements. Many existing technology alternatives can implement our model's vision. This section summarizes the technologies that our model could build on and identifies areas in which new technologies are required.

In the model spaces that operate on identified information, PI usage policies govern what companies are allowed to do with PI, for instance, specifying usage purpose, retention time and obligations. To specify the content of PI usage policies, various *policy languages* are available, some with a within-enterprise scope (e.g., EPAL), and others contemplated to be used on a web scale (e.g., PrimeLife's policy language). However, to date, no uniformly accepted global standard for electronically representing PI usage policies has emerged. The most complete standard that has been put into force, but not adopted by companies is the P3P standard published by the W3C (P3p, 2006). Harmonizing efforts based on recognized standards within the eXtensible Markup Language's framework are needed.

Negotiating PI usage policies between CR-Hs and customers is a laborious and time-consuming task. *Software agents* that can semantically understand PI usage policy content and automatically negotiate policies respecting the customer preferences are required. Many browsers' Do Not Track functionality can negotiate binary policies containing PI disclosure or secrecy (Wang and Cranor, 2011). However, usable agents that can handle the full complexity of PI usage purposes are required. The P3P standard (P3p, 2006), for instance, made a first advance in this direction.

Visibility and identity of the CR-Hs to the user can be ensured by user interface and technological means. *Standardized graphical user interface symbols* have been designed to indicate PI usage practices to users (e.g., Hansen, 2009). Also nutrition labels have been put forth (Cranor, 2012). But simple, easily understandable symbols signaling who the one visible CR-H is to users have still not been deployed. On a technical level, existing *identity assurance standards* such as SAML (Oasis, 2008) are needed to secure the authenticity of interacting agents and parties.

To manage and store PI usage policies and associated PI under user control, *policy repositories and personal data vaults* are required. Existing platforms provide repositories for limited areas, such as social networking (e.g., Lockr, Persona) or PI management (e.g., MyLifeBits). However, we lack technologies that are interoperable with any software agents and that can manage any type of PI. To ensure secure communication of PI between the entities interacting in distributed, customer-controlled PI-management environments, *advanced crypto-protocols* must be employed. For instance, fair exchange protocols could be used to guarantee that companies do not tap PI from repositories without providing in return the desired customized service to the user (Asokan et al., 1998).

To ensure accountable, policy-compliant use of PI, different *accountability management platforms* have been proposed. Sticky policies attach PI usage policies as metadata to the PI (e.g., Pearson and Mont, 2011). Trusted computing platforms controlled by the CR-H can track the enforcement of sticky PI usage policies. These policies are currently under development and often discussed under the term "trust frameworks" (Nguyen et al., 2013, p. 236). First steps towards auditability and traceability of PI use on the web were made by accountable data exchange protocols such as HTTPA (Seneviratne, 2012).

Providing a reciprocal, *open access safe harbor for big data* requires technologies that make people data redundantly and publicly available to market participants. Peer-to-peer data hosting architectures like Freenet (Clarke et al., 2001) would allow for the creation of a safe harbor for big data without revealing who contributes to and uses data from the space. For anonymizing PI at the anonymity frontier, standardized and trustworthy *anonymization technologies* are required. Privacy-preserving data mining technologies, for example, allow users to analyze large amounts of safe big data without jeopardizing its anonymity (Aggarwal and Yu, 2008).

## 6. Enforcing the four-space vision

A key challenge to the four-space vision is its enforceability. We proposed various interlocking technical and legal mechanisms to make data markets work, but these need to be globally implemented and enforced to be real bridges for safe data exchange across continents. Recent years have shown how difficult it is to reach international consensus on data protection or privacy and to enforce agreed provisions. The Safe Harbor Agreement between the U.S. and Europe on data handling practices has been quoted as an example of failure. A more effective path could be to implement and enforce binding law for data protection. To pave the way for internationally enforcing our model, we propose five building blocks based on harmonized hard law:

- technical and process standards,
- internal compliance procedures,
- external audits,
- sanctions,
- and damage claims for customers.

This five-partite enforcement strategy already proved effective for increasing consumer safety in many other areas, such as in the food and automotive industries to ensure hygiene or safety standards, for instance.

Globally binding technical and process standards for handling data in accordance with our model need to be defined. For example, specifying best-available techniques reference documents (BREFs) can provide a globally harmonized base for acceptable levels of anonymization. As a first step towards defining at least transatlantic standard bridges, the U.S. Federal Trade Commission (FTC) and the (envisaged) European Data Protection Board (EDPB) could co-operate to specify the proposed measures and then have them mandated by their respective regulatory bodies. In Europe this could be the European Commission who could do so through the delegated acts instrument foreseen in the upcoming European data protection regulation.

To ensure internal compliance and satisfy themselves with their processes handling PI, CR-Hs would need to conduct regular internal audits of their data handling processes (as is done for most business critical processes today). Auditing is facilitated by the technical accountability management platform we propose which enables reliable tracking of PI flows. Ensuring internal compliance goes in line with current calls for regular privacy impact assessments and obtaining certified privacy seals.

In the long-run it is probably necessary to have independent and external auditors check the proper handling of personal data processing and to check the operation of the accountability management platform. Similar to the provisions of the Sarbanes-Oxley Act on financial reporting, CR-Hs cannot get approval of further operating their data services if they cannot testify regular audit reports to supervisory authorities. Furthermore, CR-Hs should be obliged to notify the supervisory authorities of any irregular incidents or security breaches detected on the accountability management platform; a measure similar to the ‘data breach notifications’ that are now foreseen in the upcoming EU data protection regulation and that have been proven successful practice in many U.S. states. If there is particular suspicion of misconduct, data protection authorities or the FTC in the U.S. could conduct external audits by themselves.

If rules are not obeyed, internationally harmonized, effective sanctions should be a last remedy for enforcing the four-space model. Currently existing data protection sanctioning systems are inhomogeneous, with comprehensive, but ‘toothless’ sanctions in the EU and only punctual, but tuff penalties in the U.S. One point of reference towards a globally harmonized sanctioning systems could be the current EU data protection regulation proposal to fine infringers up to 2% of their annual worldwide turnover (Art. 79 (Ep, 2013)).

Giving customers an effective enforcement tool as well, they should have a harmonized claim for damages that result from violations against the provisions of our four-space model. Because determining the height of damages is challenging, tables of minimum damages would need to be developed as proposed by Traung (2012, p. 42). To give customers an active role in enforcing their rights, they shall have a direct right to file actions.

## 7. Discussion

Our vision for a personal information market establishes compromise between players in the current PI ecosystem and data protection proponents. Our model embraces data richness as the future of a digital economy and creates room for information-rich services and data trading as well as identified customer relationships. At the same time, our technical suggestions to help enforcing legal regulation empower people to participate in PI markets with their privacy protected. Supporting people to understand their transactions with companies and the value of their PI, we create a new and simple market structure that assigns template rights and obligations to all market players. Although this benefit applies to customers, who will have more trust in the face of market transparency, it also aids companies and legal enforcers.

Most of the legal requirements and technical enablers we propose are not new. They have been proposed for over two decades by researchers in privacy, identity and security research and debated by companies and regulators. However, no one has demonstrated how all of the puzzle pieces could be arranged in a market model to benefit both people *and* companies. No one has holistically discussed how the interaction of privacy-enhancing technologies and law can actively support a new and simple market regime, one that promises more clarity and accountability in business transactions as well as new streams of income. The model should not be misunderstood as a finalized legal and technological regulatory framework in which its components are not open for further development and extension. Also, our model is intentionally not oriented at the body of existing jurisprudence and does not discuss problems of legal integration that may ensue in particular legislations. Rather, it should be regarded as a 'social plastic' that provides a visionary route to design and change personal data markets for privacy. Most importantly, this plastic is not only originating from the legal discipline, but also covers the proposals made in computer science.

A core benefit of our model is its main technical challenge: the creation of a safe harbor for big data, a free market space that ensures anonymity. Ensuring anonymity becomes more difficult as technology becomes more powerful, facilitating identification. However, reasonable anonymity for people data will suffice if the privacy risks in case of re-identification are minimized. Sufficient anonymity can be secured if regulators enforce the anonymity requirement through rigorous sanctions for misconduct and if data protection authorities define strong and timely "BATs" (Best Available Techniques) (Directive 2010/75/EU).

A bitter pill that companies have to swallow is to finally provide people with a privacy-friendly default service option. But, as we show in this article, it isn't that a drop of bitterness. Companies can re-enter competition on the basis of service qualities. Furthermore, our model meets the privacy preferences of different individuals: Access to content at potentially lower cost for those who are willing to 'pay' with their PI and alternative versions for customers that are concerned about their data. Data protection rights proponents may argue that this preference-based market structure disadvantages the poor, who may be forced to sell their PI. This argument is true only if marketers choose to have people pay for the privacy-friendly version. Marketers could also make the data-rich version more attractive from a service perspective – with greater functionality and no ads – while offering a baseline service with non-personalized ads in an unfiltered and non-manipulated way.

To enable fair competition in markets with default privacy-friendly service versions, we suggest regulating the identified market space. One market regulation may be enforcing a price cap and a minimum service quality obligation for privacy-friendly services. Price regulation is common for many service and product areas including books, public housing, utilities, parks and roads. And even if people are asked to pay more than they do now, we argue that other services areas have seen successful transitions from an initial free offering to paid-for offerings as well; for example, the short message service (SMS) has become an important source of income for mobile operators even though it was initially a free byproduct of telephony services. Finally, even if individuals opt to use their PI in exchange for the service, we believe that our market proposal creates ample room for privacy protection: After all, companies would be accountable and liable for how they use PI. Limitless reuse and repackaging out of PI's original context would be outlawed. Privacy risks would hence be limited even for those who share. As customers will have ownership rights to their PI, they will also be brought back to the negotiating table. Ownership rights, a right to privacy-friendly service options and defaults, company accountability and a transparent market structure promise to re-establish the trust we need to see information services flourish.

The customer-controlled data space gives customers a new, self-contained way of managing their data. Our model recognizes customer-controlled data and CR-H-controlled data as equal control regimes. Customers managing their data in personal data vaults do not only have self-determined control. Personal data vaults also make customers 'feel' their ownership claim and give data into their hands. A whole new industry of trusted third parties and operators of 'personal cloud' services is evolving, enabling alternative and privacy-friendly ways of monetizing on customer data.

Finally, one more fundamental challenge of our model must be considered: the ownership claim to information. The idea that personal data could be recognized as a kind of property originated in the U.S.. This idea has been met by the criticism that people shouldn't be "propertized" (Noam, 1997; Cohen, 1999) as well as a series of other arguments (for an overview see Schwartz (2003)). Ralph Waldo Emerson once remarked: "As long as our civilization is essentially one of property, of fences, of exclusiveness, it will be mocked by delusions." For these reasons, we view the idea of legal property rights to PI critically. However, because markets already treat PI as their property, we ask only that people get the same rights that companies have already claimed for themselves and that we bring people back to the negotiating table. An ownership claim would not substitute, but rather enhance the human rights basis of privacy (Purtova, 2011). In Europe, it would provide people with better access to existing, well-proven enforcement structures. Customers could effectively claim their rights to PI on their own instead of calling on data protection authorities only. Even though data protection authorities have tried to support customers in cases of data breach, their effectiveness is limited. Most importantly, they do not have the capacity to handle the volume of cases that require settlement in personal data markets. It therefore seems more appropriate to give people access to existing market structures and tool sets.

## Acknowledgments

We want to thank so many of our colleagues that helped to challenge our evolving versions of the market model vision; in particular the 13 + 7 interview partners who took several hours to discuss all its details. We also want to thank Julian Cantella for editing this paper.